\documentstyle[a4,fleqn]{article}
\title{\bf Geometrization of perfect fluid in 5-D Kaluza-Klein theory.}
\author{\it  Sergey S. Kokarev\thanks{e-mail: sergey@yspu.yaroslavl.ru}
\\ 150000, Russia, Yaroslavl, YSPU, r.409}
\date{}

\begin{document}
\maketitle
\begin{abstract}

General formulation  of geometrization matter problem
by scalar field $\varphi =\sqrt{-G_{55}}$ with the help of possibilities
of classical \mbox{5-D} Kaluza-Klein theory is given.
Mathematical integrability conditions
for such geometrization for the case of perfect fluid are derived.
\end{abstract}

\section{\hspace{-0.7cm}.\hspace{0.5cm} Introduction}

When GR had been built and gravitational interactions
with\-in its frame had been
ge\-o\-me\-tri\-zed, Einstein himself considered
this problem as half solved:
energy-momentum tensor, that appeared in righthand part of field equations,
in his mind, would had geometrical nature too \cite{ein}.

\begin{sloppypar}
Such a possibility naturally appears whithin the frame  of multidimensional
geometrical models type of Kaluza-Klein theory, of which the first
(5-dimensional)
version has been proposed by T.Kaluza in 1921 \cite{kal,vlad1}.
Extracomponents, given by ``scalar'' sector of multidimensional metric
have been used for a number of purposes. In present article
the possibility of description of matter sources in terms of geometrical
scalar field $\varphi =\sqrt{-G_{55}}$ in effective 4-D space-time is
analyzed. In present time such an approach is worked out by
Wesson and others (see Ref. in \cite{Wes1,Wes2,Wes3,WesP}),
but some steps to the problem
have been made by the number of authors earlier \cite{shm,vlad2}.
\end{sloppypar}

\section{\hspace{-0.7cm}.\hspace{0.5cm} Statement of the problem}

    The starting point of our investigation is  vacuum 5-D Einstein
    equations, that we shall write in the form:
\begin{equation}\label{ein5}
^5R_{AB} - {1\over2}G_{AB}{}^5R = 0,
\end{equation}
where
$^5R_{AB}$ -- 5-D Ricci tensor,
$^5R$ -- 5-D curvature scalar,
$G_{AB}$ -- 5-D metric,
which in 4-D presentation has the following kind:
\begin{equation}\label{red}
G_{AB}=
\pmatrix{
\begin{tabular}{c|c}
$\tilde{g}_{\mu\nu}-(4k/c^{4})\varphi^{2}A_{\mu}A_{\nu}$&
$(2\sqrt{k}/c^{2})\varphi^{2}A_{\mu}$\\
\\
\hline
\\
$(2\sqrt{k}/c^{2})\varphi^{2}A_{\nu}$&
$-\varphi^{2}$\\
\end{tabular}
}
\end{equation}
    Expression (\ref{red}) for $G_{AB}$ can be derived with the help
    of  method of 1+4-splitting of 5-D manifold, which is outlined in
    monography \cite{vlad2}.
    ``Vector'' sector $G_{5\mu}$ is identified with
an electromagnetic  potential:
$G_{5\mu} = (2\sqrt k/c^2)\varphi^{2}A_{\mu}$,
    that lets us consider 5-D Kaluza-Klein models as unified theory of
    gravitational and electromagnetic interactions.

    Under the identification of metric of 4-D space-time sec\-ti\-on
    $\tilde g_{\mu\nu}$ with ob\-ser\-va\-ble 4-D met\-ric $g_{\mu\nu}$
    there is a  possibility of conformal transformation of the form:
\begin{equation}\label{conf}
\tilde g_{\mu\nu} = F(\varphi)g_{\mu\nu},
\end{equation}
    where
$F(\varphi)$ -- arbitrary function of scalar field.

    For identification of components of multidimensional metric
    with 4-D values it is necessary to impose the following conditions:
\begin{equation}\label{cond1}
{\partial\tilde g_{\mu\nu}\over\partial x^{5}}=0,
\qquad
{\rm\ (cylindricity\ condition)}
\end{equation}
    and restrictions on coordinate transformations:
\begin{equation}\label{coord1}
x'^{\mu}=x'^{\mu}(x^{0},x^{1},x^{2},x^{3});
\end{equation}
\begin{equation}\label{coord2}
x'^{5}= x'^{5} + f(x^{0},x^{1},x^{2},x^{3}).
\end{equation}
    Such transformations on the one hand conserve cylindricity
    conditions (\ref{cond1})
    and 4-D covariancy of $\tilde g_{\mu\nu}, A_{\mu}, \varphi$,
    and, on the other hand, are 4-D general coordinate transformations
      -- (\ref{coord1}) and gauge transformations of
      vector potentials  $A_{\mu}$ -- (\ref{coord2}).

      The method of 1+4-splitting lets us split 5-D Einstein
      equations (\ref{ein5}) in a ten 4-D Einstein equation:
\begin{equation}\label{ein4}
^4\tilde R_{\mu\nu}- 1/2\tilde g_{\mu\nu}{}^4\tilde R =
\varphi^{2}\kappa T^{(em)}_{\mu\nu}+{1\over\varphi}
(\varphi_{;\tilde\nu;\tilde\mu}-
\tilde g_{\mu\nu}\tilde{\nabla}^{2}\varphi);
\end{equation}
    four equations, corresponding to Maxwell theory:
\begin{equation}\label{max}
\tilde F^{\mu\nu}{}_{;\tilde{\nu}}-
{2\varphi_{,\alpha}\over\varphi}\tilde F^{\alpha\mu}=0
\end{equation}
    and the 15-th equation, that connects invariants of gravitational
    and electromagnetic fields:
\begin{equation}\label{scal}
^{4}\tilde R+{3\over8\pi}\kappa\varphi^{2}
\tilde F_{\alpha\beta}\tilde F^{\alpha\beta} =0.
\end{equation}
In  (\ref{ein4})---(\ref{scal}) values with ``tilde'' are related
to starting metric
$\tilde g_{\mu\nu}$.

        In present article all investigations are car\-ried out, at first,
        with\-out an ele\-ctro\-mag\-ne\-tic field ($F_{\mu\nu} = 0$),
        and secondly, conformal factor in (\ref{conf})
        is taken in the form:
\begin{equation}\label{conf1}
F(\varphi)= \varphi^{2n},
\end{equation}
    where $n$ --  arbitrary real constant.
    Then  equations (\ref{ein4}) and (\ref{scal}) with
    account of  conformal
    transformation (\ref{conf1}) can be put to the form:
\begin{equation}\label{ein4c}
^{4}R_{\mu\nu}-{1\over2}g_{\mu\nu}{}^{4}R=
(1+2n)\phi_{;\mu;\nu}-(2n^{2}+2n-1)\phi_{,\mu}\phi_{,\nu}-
\end{equation}
\[
-g_{\mu\nu}((1+2n)\nabla^{2}\phi + (n^{2}+n+1)(\nabla\phi)^{2});
\]

\vspace{0.2cm}
\begin{equation}\label{scalc}
n\nabla^{2}\phi+ n^{2}(\nabla\phi)^{2}- 1/6{}^{4}R=0,
\end{equation}
    where
$\phi =\ln\varphi.$
Equations (\ref{max}) are satisfied identically.

       So, when all outlined above conditions are satisfied, then vacuum
       Einstein equations (\ref{ein5})
      (after 1+4-splitting procedure) take the form of nonvacuum 4-D
       Einstein equations (\ref{ein4c}), (\ref{scalc}).

       Let us assume now, that we have some exact solutions of 4-D
       nonvacuum Einstein equations with energy-momentum tensor
       of perfect fluid in right-hand side($\kappa = 1$):
\begin{equation}\label{gr}
R_{\mu\nu}-1/2g_{\mu\nu}R=
(p+\varepsilon)u_{\mu}u_{\nu}-pg_{\mu\nu}\equiv T_{\mu\nu}^{(hd)}
\end{equation}
\begin{sloppypar}
    {\it If scalar field $\phi$ and constant  $n$ are so, that
\begin{equation}\label{geom}
(p+\varepsilon)u_{\mu}u_{\nu}-pg_{\mu\nu}=
\end{equation}
\[(1+2n)\phi_{;\mu;\nu}-
(2n^{2}+2n-1)\phi_{\mu}\phi_{\nu}-g_{\mu\nu}((1+2n)\nabla^{2}\phi+(n^{2}+n+1)(\nabla\phi)^{2}).
\]
 and equation  (\ref{scalc}) is satisfied,  then we can say, that
such  $T_{\mu\nu}^{(hd)}$ can be geometrized within the frame of
5-D Kaluza-Klein theory and 4-D metric, generated by this sources,
has a purely 5-D origin.}
    Below we shall derive restrictions on the geometry of 4-D space-time and
    movement of matter there, under which the equation (\ref{geom})
    is satisfied.
\end{sloppypar}

\section{\hspace{-0.7cm}.\hspace{0.5cm} Integrability conditions}

    Equations (\ref{geom}) can be considered as a system of differential
    equations in partial derivatives of the second order for components
    of the gradient $\phi_{,\mu}$, taking $u_{\mu}, \varepsilon, p, g_{\mu\nu}$
    as prescribed functions.
    For integrability analysis of this system  it is convenient to
    rewrite it in following more general form:
\begin{equation}\label{int1}
\phi_{;\mu;\nu} = k\phi_{\mu}\phi_{\nu}+au_{\mu}u_{\nu}+bg_{\mu\nu},
\end{equation}
where:
\[
k=-{k_{2}\over k_{1}};\ \ \ \ a={p+\varepsilon\over k_{1}};
\]
\begin{equation}\label{abb}
b={-p-k_{3}(\nabla\phi)^{2}\over k_{1}}=c+\overline k(\nabla\phi)^{2};
\end{equation}
\[
c=-{p\over k_{1}};\ \ \ \ \overline k=-{k_{3}\over k_{1}};
\]
\[
k_{1}=1+2n\not=0;\ \ \ \ k_{2}= -2n^{2}-2n+1;\ \ \ \  k_{3}= 3n^{2}+3n
\]

The equation (\ref{int1}) can be derived from (\ref{geom}) with using the relation:
\begin{equation}
\nabla^{2}\phi=-(1+2n)(\nabla\phi)^{2},
\end{equation}
    which, in turn, follows from linear combination of
equation (\ref{scalc}) and contracted with $g_{\mu\nu}$ equation (\ref{ein4c}).

    Integrability conditions for system (\ref{int}) imply identical
    fulfillment of the relationship
\begin{equation}\label{int}
2\phi_{;\mu;[\nu;\lambda]}=R^{\sigma}_{\cdot\mu\nu\lambda}\phi_{\sigma},
\end{equation}
    if into the left-hand side expressions from right-hand side
    of (\ref{int1}) are inserted  ,
    with replacement of second covariant derivatives  $\phi_{;\mu;\nu}$
    through (\ref{int1}) again.
    Identical fulfillment  (\ref{int}) is necessary for the existence
    of a scalar field  $\phi$,  satisfied the equation~(\ref{geom}).

       Omitting intermediate calculations, we perform (\ref{int}) in the
        form:
\begin{equation}\label{int2}
R^{\sigma}_{\cdot\mu\nu\lambda}\phi_{\sigma}=
u_{\mu}(u_{\nu}a_{\lambda}-u_{\lambda}a_{\nu})+
kau_{\mu}(u_{\lambda}\phi_{\nu}-u_{\nu}\phi_{\lambda})+
(c_{\lambda}g_{\mu\nu}-c_{\nu}g_{\mu\lambda})+
\end{equation}
\[
\xi_{1}(\phi_{\lambda}g_{\mu\nu}-\phi_{\nu}g_{\mu\lambda})+
\xi_{2}(u_{\lambda}g_{\mu\nu}-u_{\nu}g_{\mu\lambda})+
a(u_{\mu;\lambda}u_{\nu}-u_{\mu;\nu}u_{\lambda}+
u_{\mu}(u_{\nu;\lambda}-u_{\lambda;\nu})),
\]

\vspace{0.2cm}
\begin{equation}\label{abb1}
{\rm where}\ \
\xi_{1}= 2\overline k(k(\nabla\phi)^{2}+b)-kb;\ \
\xi_{2}=2\overline ka(\vec{\nabla}\phi\cdot\vec u)
\end{equation}

\section{\hspace{-0.7cm}.\hspace{0.5cm} Tetrads formalism}

       For obtaining of invariant form of integrability condition (\ref{int1})
       it is
       convenient to go to tetrad component of all tensors.
As a tetrad basis we take Lorentz tetrad
$\vec t, \vec x, \vec y, \vec z$,
The first of noted vectors can be identified with
4-velocity of matter $\vec u$:
\begin{equation}\label{tetr}
g=\vec u\otimes\vec u-\vec x\otimes\vec x-\vec y\otimes\vec y-
\vec z\otimes\vec z.
\end{equation}
    Curvature tensor
$R_{\mu\nu\lambda\sigma}$
is represented in such basis in the following kind:
\begin{equation}\label{curv}
R_{\mu\nu\lambda\sigma}=a_{ij}X_{\mu\nu}^{(i)}X_{\lambda\sigma}^{(j)}\ \
i,j=\overline{1,6},
\end{equation}
where
$X_{\mu\nu}^{(i)}$ -- simple bivectors,
that forms the basis of 6-dimensional
linear bivector space and are built from basic tetrad vectors:
\begin{equation}\label{biv}
\begin{array}{ll}
X_{\mu\nu}^{(1)}=2u_{[\mu}x_{\nu]}; &
X_{\mu\nu}^{(2)}=2u_{[\mu}y_{\nu]};\\
X_{\mu\nu}^{(3)}=2u_{[\mu}z_{\nu]};  &
X_{\mu\nu}^{(4)}=2x_{[\mu}y_{\nu]};\\
X_{\mu\nu}^{(5)}=2z_{[\mu}x_{\nu]}; &
X_{\mu\nu}^{(6)}=2y_{[\mu}z_{\nu]}.
\end{array}
\end{equation}
    Matrix $a_{ij}$ -- is symmetrical and its elements are tetrad components
    of curvature tensor, for example:
\begin{equation}\label{curvt}
R_{\mu\nu\lambda\sigma}u^{\mu}x^{\nu}u^{\lambda}x^{\sigma}=
R_{(0)(1)(0)(1)}=a_{11}\ \ {\rm ¨\ â. ¤.}\dots
\end{equation}
    Einstein equations (\ref{gr}) in tetrad form take the following kind:
\begin{equation}\label{grt}
\begin{array}{lll}
 a_{35}=a_{24};& a_{14}=a_{36};& a_{26}=a_{15};\\
a_{12}=-a_{56};& a_{13}=-a_{46};& a_{23}=-a_{45};\\
\end{array}
\end{equation}
\[
\begin{array}{ll}
-a_{44}-a_{55}-a_{66}=\varepsilon;& a_{66}-a_{22}-a_{33}=p;\\
a_{55}-a_{11}-a_{33}=p;& a_{44}-a_{11}-a_{22}=p.\\
\end{array}
\]

In the right-hand side  of (\ref{int1})  all tensors can be decomposed by a
tetrad basis as  follows:
\begin{equation}
\vec{\nabla}\phi= \phi_{0}\vec u+\phi_{1}\vec x+\phi_{2}\vec y+\phi_{3}\vec z;
\end{equation}
\begin{equation}
\vec{\nabla}\varepsilon=\varepsilon_{0}\vec u+\varepsilon_{1}\vec x+
\varepsilon_{2}\vec y+\varepsilon_{3}\vec z;
\end{equation}
\begin{equation}
\vec{\nabla}a={1+p_{\varepsilon}\over k_{1}}\vec{\varepsilon};
\quad
\vec{\nabla}c=-{p_{\varepsilon}\over k_{1}}\vec{\nabla}\varepsilon.
\end{equation}
    Here
$p_{\varepsilon} = dp/d\varepsilon$.

For decomposition of the covariant derivative of vector $u_{\mu}$
we use the well known expression \cite{vlad2}:
\begin{equation}
u_{;\mu;\nu}=F_{\mu}u_{\nu}+\omega_{\mu\nu}+
\sigma_{\mu\nu}-{\theta\over3}h_{\mu\nu},
\end{equation}
    where
\[
F_{\mu} = u_{\mu;\nu}u^{\nu}
\]
 --  the acceleration vector of a comoving reference frame,
$F_{\nu}u^{\nu} = 0$;
\begin{equation}
\omega_{\mu\nu} = u_{[\mu;\nu]} + u_{[\mu}F_{\nu]}
\end{equation}
-- the antisymmetric tensor of angular velocity of the reference frame,
$\omega_{\mu\nu}u^{\nu} = 0$;
\begin{equation}
\sigma_{\mu\nu} = u_{(\mu;\nu)} - u_{(\mu}F_{\nu)}+{\theta\over3}h_{\mu\nu}
\end{equation}
-- the symmetric shear tensor (traceless part of strain tensor)
  of the reference frame,
$\sigma_{\mu\nu}u^{\nu} = 0$;
\begin{equation}
\theta = u^{\mu}{}_{;\mu};
\end{equation}
-- the stretch scalar of the reference frame (trace of strain tensor);
\begin{equation}
h_{\mu\nu} = u_{\mu}u_{\nu} - g_{\mu\nu}
\end{equation}
-- the metric of the local 3-D space section,
orthogonal to $u_{\mu}$.

Tensors of the reference frame $F, \omega $ ¨ $\sigma $
in a chosen tetrad basis are given by following expressions:
\begin{equation}
\vec F=F_{1}\vec x+F_{2}\vec y+F_{3}\vec z;
\end{equation}
\begin{equation}
\omega_{\mu\nu}=\omega_{4}X_{\mu\nu}^{(4)}+\omega_{5}X_{\mu\nu}^{(5)}+
\omega_{6}X_{\mu\nu}^{(6)};
\end{equation}
\begin{equation}
\sigma_{\mu\nu}=\sigma_{2}Y_{\mu\nu}^{(2)}+\sigma_{3}Y_{\mu\nu}^{(3)}-
(\sigma_{2}+\sigma_{3})Y_{\mu\nu}^{(4)}+\sigma_{8}Y_{\mu\nu}^{(8)}+
\sigma_{9}Y_{\mu\nu}^{(9)}+\sigma_{10}Y_{\mu\nu}^{(10)}
\end{equation}
    In the last expression
$Y_{\mu\nu}^{(i)}$
-- diadic basis of 10-dimensional linear space of symmetric tensors,
formed by basic vectors:
\begin{equation}
\begin{array}{ll}
Y_{\mu\nu}^{(1)}=u_{\mu}u_{\nu}; & Y_{\mu\nu}^{(2)}=x_{\mu}x_{\nu};\\
Y_{\mu\nu}^{(3)}=y_{\mu}y_{\nu}; & Y_{\mu\nu}^{(4)}=z_{\mu}z_{\nu};\\
Y_{\mu\nu}^{(5)}=2u_{(\mu}x_{\nu)}; & Y_{\mu\nu}^{(6)}=2u_{(\mu}y_{\nu)};\\
Y_{\mu\nu}^{(7)}=2u_{(\mu}z_{\nu)}; & Y_{\mu\nu}^{(8)}=2x_{(\mu}y_{\nu)};\\
Y_{\mu\nu}^{(9)}=2x_{(\mu}z_{\nu)}; & Y_{\mu\nu}^{(9)}=2y_{(\mu} z_{\nu)}.
\end{array}
\end{equation}
  The  tensor $h_{\mu\nu}$ has  the following form:
\begin{equation}
h=\vec x\otimes\vec x+\vec y\otimes\vec y+\vec z\otimes\vec z.
\end{equation}

    Integrability conditions (\ref{int1}),  being rewritten in tetrad form
    are invariant under 4-D general coordinates transformations.
    Freedom in the choosing of tetrad basic vectors is reduced to
    local space rotations of triad $\vec x, \vec y,\vec z$ ,
    and can be used for the simplification
    of the  obtained in the next section integrability equations.
    For the most generality we suppose that triad $\vec x, \vec y, \vec z$
    is arbitrary oriented.

\section{\hspace{-0.7cm}.\hspace{0.5cm} Integrability conditions
in tetrads representation.}

Equating scalar coefficient in (\ref{int1}) under the similar combinations of
basic vectors
in right-hand and in the left-hand sides we get the following system
of scalar equations, that is equivalent to the original system (\ref{int1}):

\[
a_{11}\phi_{1}+a_{12}\phi_{2}+a_{13}\phi_{3}=e_{1}+\lambda\phi_{1}-f_{1};
\]
\[
a_{12}\phi_{1}+a_{22}\phi_{2}+a_{23}\phi_{3}=e_{2}+\lambda\phi_{2}-f_{2};
\]
\[
a_{13}\phi_{1}+a_{23}\phi_{2}+a_{33}\phi_{3}=e_{3}+\lambda\phi_{3}-f_{3};
\]
\[
a_{14}\phi_{1}+a_{24}\phi_{2}+a_{34}\phi_{3}=2W_{4};
\]
\[
a_{15}\phi_{1}+a_{25}\phi_{2}+a_{35}\phi_{3}=2W_{5};
\]
\[
a_{16}\phi_{1}+a_{26}\phi_{2}+a_{36}\phi_{3}=2W_{6};
\]
\[
a_{11}\phi_{0}+a_{14}\phi_{2}-a_{15}\phi_{3}=\Delta_{2}+S;
\]
\[
a_{12}\phi_{0}+a_{24}\phi_{2}-a_{25}\phi_{3}=\Delta_{8}+W_{4};
\]
\[
a_{13}\phi_{0}+a_{34}\phi_{2}-a_{35}\phi_{3}=\Delta_{9}-W_{5};
\]
\[
a_{14}\phi_{0}+a_{44}\phi_{2}-a_{45}\phi_{3}=
p_{\varepsilon}e_{2}-\xi_{1}\phi_{2};
\]
\[
a_{15}\phi_{0}+a_{45}\phi_{2}-a_{55}\phi_{3}=
-p_{\varepsilon}e_{3}+\xi_{1}\phi_{3};
\]
\[
a_{16}\phi_{0}+a_{46}\phi_{2}-a_{56}\phi_{3}=0;
\]
\[
a_{12}\phi_{0}-a_{14}\phi_{1}+a_{16}\phi_{3}=\Delta_{8}-W_{4};
\]
\[
a_{22}\phi_{0}-a_{24}\phi_{1}+a_{26}\phi_{3}=\Delta_{3}+S;
\]
\[
a_{23}\phi_{0}-a_{34}\phi_{1}+a_{36}\phi_{3}=\Delta_{10}+W_{6};
\]
\[
a_{24}\phi_{0}-a_{44}\phi_{1}+a_{46}\phi_{3}=
-p_{\varepsilon}e_{1}+\xi_{1}\phi_{1};
\]
\[
a_{25}\phi_{0}-a_{45}\phi_{1}+a_{56}\phi_{3}=0;
\]
\[
a_{26}\phi_{0}-a_{46}\phi_{1}+a_{66}\phi_{3}=
p_{\varepsilon}e_{3}-\xi_{1}\phi_{3};
\]
\[
a_{13}\phi_{0}+a_{15}\phi_{1}-a_{16}\phi_{2}=\Delta_{9}+W_{5};
\]
\[
a_{23}\phi_{0}+a_{25}\phi_{1}-a_{26}\phi_{2}=\Delta_{10}-W_{6};
\]
\[
a_{33}\phi_{0}+a_{35}\phi_{1}-a_{36}\phi_{2}=\Delta_{4}+S;
\]
\[
a_{34}\phi_{0}+a_{45}\phi_{1}-a_{46}\phi_{2}=0;
\]
\[
a_{35}\phi_{0}+a_{55}\phi_{1}-a_{56}\phi_{2}=
p_{\varepsilon}e_{1}-\xi_{1}\phi_{1};
\]
\[
a_{36}\phi_{0}+a_{56}\phi_{1}-a_{66}\phi_{2}=
-p_{\varepsilon}e_{2}+\xi_{1}\phi_{2}.
\]
Here
\begin{equation}\label{abb2}
e_{i}=\varepsilon_{i}/k_{1};\ \ f_{i}=aF_{i};\ \ W_{i}=aw_{i};\ \
\Delta_{i}=a\sigma_{i};\ \ \lambda=\xi_{1}-ka.
\end{equation}
\[
S={-p_{\varepsilon}\over k_{1}}\varepsilon_{0}+\xi_{1}\phi_{0}+
\xi_{2}-a\theta/3.
\]

     The obtained equations connect  components of a curvature tensor
     of 4-D space with a characteristics of motion of a matter there.
     Besides, this connections are additional to Einstein equations (\ref{gr}).
     If space-time allows the identical fulfillment of the obtained
     integrability
     equations
     and scalar field satisfy the equation (\ref{scalc}), then
     this space-time has 5-D nature and 5-dimensionally
     geometrized matter.

\section{\hspace{-0.7cm}.\hspace{0.5cm} Example: potential motion in the
Friedman flat cosmological model.}\label{example}

       Let us suppose, that expansion $\phi_{,\mu}$
       over basic tetrad vectors has the following form:
\begin{equation}
\phi_{,\mu}=\phi_{0}u_{\mu}.
\end{equation}
    This means, that a vector field  $u_{\mu}$ is orthogonal to
    hypersurface $\phi = const$. For simplicity we restrict our
    attention to the case of motion in the Friedman flat cosmological
    model:
\begin{equation}\label{frid}
ds^{2}=dt^{2}-e^{2\lambda}(dx^{2}+dy^{2}+dz^{2}).
\end{equation}
    Nonzero components of curvature tensor are:
\begin{equation}
a_{11}=a_{22}=a_{33}=\ddot{\lambda}+\dot{\lambda}^{2},\ \
-a_{44}=-a_{55}=-a_{66}=\dot{\lambda}^{2}.
\end{equation}
    Tetrad Einstein equations (\ref{grt}) take the following form:
\begin{equation}\label{eincosm}
3\dot{\lambda}^{2}=\varepsilon;\ \  -2\ddot{\lambda}-3\dot{\lambda}^{2}=p.
\end{equation}
  4-velocity $u_{\mu}$ has the following components:
\begin{equation}
u_{0}=1,\ \  u_{1}=u_{2}=u_{3}=0.
\end{equation}
    From the tensors characteristics of the reference frame
    only stretch coefficient is nonzero:
\begin{equation}
\theta=3\dot{\lambda}.
\end{equation}

         Integrability conditions of the previous section  are reduced to only:
\begin{equation}\label{intc}
a_{11}\phi_{0}=S.
\end{equation}
    Taking into account that $\phi_{0} = \dot\phi$,  writing out abbreviated
    designations by formulae  (\ref{abb1}) and (\ref{abb2}) and expressing
     $\dot\varepsilon$ from Einstein equations (\ref{eincosm}),
     condition (\ref{intc}) can be put to the form:
\begin{equation}\label{intred}
\mu_{1}\dot{\phi}^{3}+\dot{\phi}(\dot{\lambda}^{2}(\mu_{2}-1)-
\ddot{\lambda})-\mu_{3}\dot{\lambda}\ddot{\lambda}=0,
\end{equation}
    where
\[
\mu_{1} = \overline k(k + 2\overline k);\ \
\mu_{2} = 3(2\overline k/k_{1} +p_{\varepsilon}k/k_{1});\ \
\mu_{3} =2(3p_{\varepsilon} - 1)/k_{1}
\]
    and state equation is taken in the following form:
\begin{equation}\label{state}
p=p_{\varepsilon}\varepsilon,\ \  p_{\varepsilon}=const.
\end{equation}

         Equation (\ref{scalc}), that in metric (\ref{frid}) has the form:
\begin{equation}\label{scalcc}
\ddot{\phi}+3\dot{\lambda}\dot{\phi}+(1+2n)\dot{\phi}^{2}=0
\end{equation}
  has the following first integral:
\begin{equation}\label{dotphi}
\dot{\phi}={e^{-3\lambda}\over\overline C+(1+2n)\int e^{-3\lambda}dt},
\end{equation}
    where $\overline C$ -- is a constant integration, of which
    general solution can be easily obtained:
\begin{equation}\label{phi}
\phi={1\over1+2n}\ln(\overline C+(1+2n)\int e^{-3\lambda}dt)+C_{1}
\end{equation}
    With the substitution of $\dot\phi$ from  (\ref{dotphi})
    to (\ref{scalcc})
    the expression in the left-hand must be equal zero identically.

 Einstein equations for a chosen class state equation (\ref{state}) can be
 easi\-ly in\-teg\-ra\-ted:
\[
\lambda=
\left\{\begin{array}{lr}
(2/3(1+p_{\varepsilon}))
\ln[(3/2)(1+p_{\varepsilon})t
+\overline{C}_{0}], & p_{\varepsilon}>-1;\\
C_{0}t, & p_{\varepsilon}=-1.\end{array}\right.
\]
    For  $p_{\varepsilon} = -1\ (p + \varepsilon= 0)$,
    omitting all intermediate calculations, we obtain
    that (\ref{intc}) identically
    satisfied under $n = 1,-2$. In this case $\phi = C_{0}t$.
    Case $p_{\varepsilon} = 1\linebreak (p = \varepsilon)$
    will be considered in subsection (3) of Conclusion.
    For $p_{\varepsilon} \not= \pm1$
we obtain the following square equation for the index of conformal transformation $n$:
\begin{equation}
\alpha_{1}n^{2}+\alpha_{1}n+\alpha_{2}=0,
\end{equation}
    where
\begin{equation}
\alpha_{1} = 27p_{\varepsilon}^{3}+63p_{\varepsilon}^{2} +
33p_{\varepsilon} + 5;
\end{equation}
\[
\alpha_{2} = 18p_{\varepsilon}^{2} +24p_{\varepsilon} - 10.
\]
    Its solution is:
\begin{equation}
n={3p_{\varepsilon}(\sigma-1)-(3\sigma+1)\over2(3p_{\varepsilon}+1)},\ \
\sigma=\pm1
\end{equation}
    For dust  ($p_{\varepsilon}  = 0$) we have $n = 1,\ -2$,
    for radiation $(p_{\varepsilon} = 1/3) \  n = 0,\ -1.$

      Corresponding 5-D vacuum metrics that in all cases
      can be obtained by the inverse conformal transformation are
      only of the following two types:\\
\begin{equation}\label{1}
 dI_{1}^{2}=dt^{2}-dx^{2}-dy^{2}-dz^{2}-t^{2}(dx^{5})^{2};
\end{equation}
\begin{equation}\label{2}
 dI_{2}^{2}=dt^{2}-t(dx^{2}+dy^{2}+dz^{2})-
{\displaystyle {1\over t}}(dx^{5})^{2}
\end{equation}
    Its relations with considered 4-D metrics are shown in the table:

    \vspace{0.5cm}

\begin{center}
\begin{tabular}{|c|c|c|c|}
\hline
&$p_{\varepsilon}=-1$&$p_{\varepsilon}=0$&$p_{\varepsilon}=1/3$\\
\hline
(1)&$n=1$&$n=-2$&$n=-1$\\
\hline
(2)&$n=-2$&$n=1$&$n=0$\\
\hline
\end{tabular}
\end{center}

\vspace{0.5cm}
Cases with  $n = 1$ have been considered in \cite{Kok1,Kok2}, the case with
$n = 0$ have been analyzed in Wesson's work \cite{Wes1} .

\section{\hspace{-0.7cm}.\hspace{0.5cm} Conclusion}

     Completing our account we make some remarks:
\begin{itemize}

\item[1.] The approach, proposed here, can be called  ``4-dimensional'',
because 4-D metric and an energy-momentum tensor are given.
    From integrability conditions scalar field $\phi$ and index $n$
    can be found, or, that is the same, can be found
    5-D vacuum space-time which geometrizes beforehand given matter.
     As an example of another approach, ``5-dimensional'', we can consider the
     approach of Wesson, where known 5-D vacuum solution is used.
      5-D Einstein equations are split in to parts:
      one -- 4-D Einstein tensor,  other --
    a combination of derivatives of the scalar field, which is declared as an
    effective energy-momentum tensor of induced matter.
    Type of this tensor in such approach is, in general,  arbitrary.
    In  both \cite{Wes2,Wes3} obtained energy-momentum tensor was anisotropic.
     All investigations in Wesson's  group are carried out under  $n=0$.
     Note, that using of the conformal transformation alleviates statement
     made in \cite{WesP}: there is no rigid necessity to introduce
     the 5-th coordinate into metric to obtain state equation for
     effective matter other than radiationlike.

\item[2.] The problem of the geometrization of matter in Kaluza-Klein theory
has many formal analogies with the known problem of isometric embedding
of 4-D Riemannian space into flat space of more dimensions \cite{eiz}.

\item[3.] The restriction on index $n \not = -1/2$
    involves the fact that in this case the second derivatives of a scalar
    field in equations (\ref{geom})  vanish.  Equations (\ref{geom}) became
    algebraic with respect to the gradient $\phi_{\mu}$ and integrability
    conditions will be
$$
\phi_{\mu,\nu}-\phi_{\nu,\mu}=0,
$$
    where $\phi_{\mu}$ is algebraically expressed through $u_{\mu},\ p$
    and $\varepsilon$. In the considered example a peculiar case
    $p_{\varepsilon} = 1\ \ (p = \varepsilon)$ is realized
    namely under  $n =-1/2$ from both 5-D metrics (\ref{1},\ref{2}).

\item[4.] Cosmological models of open and closed types have been considered
under  $n=1$ in \cite{vlad2}(p.230-234) from the viewpoint of ``5-dimensional''
approach.
\end{itemize}


\begin{thebibliography}{99}

\bibitem{ein}
A. Einstein, {\it  Physic and reality, } M. "Nauka" , p.159 (1974) (In Russian)

\bibitem{kal}
T. Kaluza, {\it Sitzungsber. d. Berl.Akad.}, 966-971 (1921).

\bibitem{vlad1}
Yu. S. Vladimirov, {\it Space-time: explicit and hidden dimensions,} M.Nauka,
1989 (In Russian)

\bibitem{shm}
E. Schmutzer,  {\it Exp. Techn, Phys.,} {\bf 28} pp.395-402, 499-508 (1980);
{\bf 29} pp.129-136, 337-341, 463-480 (1981).

\bibitem{vlad2}
Yu. S. Vladimirov {\it Reference frames in gravitation theory,} M.Energoizdat,
1982 (In Russian).

\bibitem{eiz}
L. P. Eisenhart, {\it Riemannian geometry,}  Izd.in.lit.,1948 (In Russian).

\bibitem{Wes1}
P. S. Wesson,  {\it Astroph. Journal,} {\bf 394} p.19-24 (1992).

\bibitem{Wes2}
P. S. Wesson, {\it Astroph. Journal,} {\bf 420} L49-L52 (1994).

\bibitem{Wes3}
P. S. Wesson, {\it Phys.Let.,} {\bf B 276} 299-302 (1992).

\bibitem{WesP}
P. S. Wesson, J. Ponce de Leon, P. Lim, H. Liu, {\it Int.J.of Mod.Phys.} {\bf D2} 163-170 (1993).

\bibitem{Kok1}
S. S. Kokarev, {\it Izv. VUZov (Physics),}  N1, p.111-117 (1995).(In Russian)

\bibitem{Kok2}
S. S. Kokarev, {\it In abstracts of shool-ceminar
"Multid. gravity and cosmology",} Yaroslavl,1994.,Œ. 1994.,p.19.
(In Russian).
 \end{thebibliography}
\end{document}